\documentclass[preprint,12pt]{elsarticle}

\usepackage{amssymb}
\usepackage{amsmath}
\usepackage{mathtools}

\usepackage{multirow}
\usepackage{braket}
\usepackage{caption}
\usepackage{subcaption}
\DeclareMathOperator\erfc{erfc}
\DeclareMathOperator\sign{sign}

\usepackage{xcolor}
\journal{\textcolor{white}{O}}

\begin{document}

\begin{frontmatter}



\title{High-dimensional autocompensating discrete modulation CV-QKD protocol  in optical fibers} %% Article title


\author{Alexandre Vázquez-Martínez$\,^{a}$, Xesús Prieto-Blanco$\,^{a}$, Eduardo F. Mateo$\,^{b}$ and Jesús Liñares$\,^{a}$} %% Author name

\affiliation{organization={Quantum Materials and Photonics Research Group, Optics Area, Department of Applied Physics, iMATUS/Faculty of Physics/Faculty of Optics and Optometry,  Universidade de Santiago de Compostela},%Department and Organization
            addressline={Campus Vida s/n}, 
             postcode={E-15782}, 
            city={Santiago de Compostela},
                       state={Galicia},
            country={Spain}}

 \affiliation[label2]{organization={Submarine Network Division, NEC Corporation },%Department and Organization
            %addressline={}, 
             postcode={108-800 1}, 
            city={Tokyo},
            country={Japan}}

%% Abstract
\begin{abstract}
In this work we present a  high-dimensional discrete modulated  CV-QKD  protocol, with dimension $2^{N}$, in optical fibers, where $N$ is  the number of optical modes. We use $N$-dimensional product states  of weak coherent states that undergo  perturbations along optical fibers and that are cancelled in a passive way (autocompensation)  and  then measured by a standard balanced homodyne detection. We analyze the security  of this high-dimensional protocol when an intercept and resend attack is implemented with simultaneous homodyne measurement  in both quadratures and the resulting secure key gains are calculated showing a significative  increase  with dimension.
\end{abstract}



%% Keywords

\end{frontmatter}




%% Labels are used to cross-reference an item using \ref command.

%%%%%%%%%%%%%%%%%%%%%%%%%%%%%%%%%%%%%
%%%%%%%%%%%%%%%%%%%%%%%%%%%%%%%%%%%%%%

\section{Introduction}\label{secintro}
Quantum Key Distribution (QKD) is considered as a key technology for future secure communications. In fact, the security of the present classical cryptography, based on mathematical algoritms, will no longer be guaranteed when quantum computers become available. QKD, a branch of quantum communications,  is a  technology based on quantum optics that allows  the exchange of secure information between two users, Alice and Bob, sharing   a random bit series or key, which can not be cloned by an eavesdropper. There are several QKD schemes which can be classified into Discrete Variable (DV) and Continuous Variable (CV) QKD  \cite{RevGisin,Pirandola2020}. CV-QKD is based on the measurement of the quadratures of the optical field by using, for example, a balanced homodyne detection, which is a standard detection method in classical coherent communications such as optical fiber communications, and accordingly CV-QKD shows an important appeal  to be used in standard optical fiber networks.

There are several types of CV-QKD protocols, one of them is the so-called Discrete Modulation   QKD (DM-CV-QKD) of coherent states which was originally proposed by Ralph \cite{Ralph1999}, Hillery \cite{Hillery2000} and Reid \cite{Reid2000} and the idea was followed by  Namiki  and Hirano \cite{Namiki2003, Namiki2006}, and Heid and Lütkenhaus \cite{Heid2006}. At present,  DM-CV-QKD protocols continue to be developed in both optical fibers and free space and present some advantages with respect to protocols based on continuous modulation, that is, DM-CV-QKD protocols simplify both the modulation scheme and  key extraction process, with respect to the continuous modulation, as for example the Gaussian-modulated CV-QKD protocols, where the key has to be extracted from a continuous random  distribution \cite{Djordjevic2019}. Likewise, DM-CV-QKD protocols are noted for long-distance applicability even at low SNR \cite{Pan2022, Rani2023}.

In this work we present a High-dimensional   (HD)   DM-CV-QKD protocol  in optical fibers by using a number  $N$  of   spatial  optical modes. It must be stressed that such a HD protocol is fully compatible with modern optical netwotks based on space division multiplexing \cite{Edu2024,Winzer2023}. In particular, we  use $N$-dimensional product states  of weak coherent states which can be  compensated in a passive way, that is,  the polarization perturbations due to the optical  fiber propagation can be removed, and therefore a proper  measurement by balanced homodyne detection can be made. These states are excited randomly either all in the first quadrature or all in the second quadrature of the optical field and next they are measured in one of the two quadratures (or {\em{basis}}) by balanced homodyne detection in each optical mode. We will have  $2^{N}$ states  in each $N$-dimensional quadrature, therefore, a HD-DM-CV-QKD protocol will be implemented, achieving thus a larger security and a larger secure key rate. Furthermore, we will take advantage of  the polarization of each spatial mode to send the quantum state (signal) in  a mode of linear polarization and the Local Oscillator (LO) in the orthogonal polarization mode and thus to implement  both  autocompesation and homodyne detection in $N$ modes.  As a clarification,  spatial channels are not coupled, and importantly the protocol is independent of the relative phase between the weak coherent states. We must stress, on the one hand, that HD Discrete Variable QKD protocols have been proposed by using multimode optical fibers \cite{Canas2017,Ding2017,Bal19}, and on the other hand, product states with only two photons have been also proposed in Discrete Variable QKD protocols as for example the MDI-QKD one \cite{CURTY}.

The plan of the paper is as follows. In Sec.\,2 we   introduce the fundamental aspects of the  HD-DM-CV-QKD protocol, that is,  we present the $N$-dimensional weak coherent quantum states along with  the main steps to implement the   protocol with  $2^{N+1}$  states, that is, $2^{N}$ states in each quadrature. In Sec.\,3  the $2^{N}$-dimensional postselection efficiency (PE) and the intrinsic quantum bit error rate (IQBER) are calculated. In Sec.\,4 the autocompensating QKD (plug and play) system is presented.  In Sec.\,5 the security of this protocol is analysed by studying  an intercept and resend attack implemented by a simultaneous double homodyne measurement. In Sec.\,6 the quantum bit error rate due to the presence of Eve is obtained and next the secure key rate gain is calculated and analysed for $N=1,2,3$ modes. In Sec.\,7 conclusions are presented.

%%%%%%%%%%%%%%%%%%%%%%%%%%%%%%%%%%%%
%%%%%%%%%%%%%%%%%%%%%%%%%%%%%%%%%%%%

\section{$2^{N}$-dimensional Discrete Modulated CV-QKD protocol }\label{sec2}

We present a $2^{N}$-dimensional   HD-DM-CV-QKD protocol which overcomes the 2-dimensional  DM-CV-QKD protocol with $N$ channels. In fact, as it will be shown, a remarkable increase in security  is achieved due to the implementation of  higher dimensionality instead of using  $N$ channels with CV-QKD that,  although they increase the key rate, they  do not improve the security performance of the protocols.  This   HD-DM-CV-QKD protocol uses multimode coherent states, in particular,  $N$-dimensional product weak coherent states $\vert \Psi\rangle=\vert \alpha_{1} ... \alpha_{N}\rangle=\ket{a_{1}\exp({i\phi_{1}}),..,a_{N}\exp({i\phi_{N}})}$, where  $a_{k}=|\alpha_{k}|=\sqrt{n_{k}} \in \mathbb{R},  \,k=1,...,N$, with  $n_{k}$ the mean photon number (which will be close to unity) and $\phi_{k}$ the phase of the coherent state.  These product coherent states can be excited in  a multimode optical fiber with $N$ modes, for example the weak coherent quantum states (signal) can be excited in $N$ spatial modes  with the same polarization,  and the orthogonal polarization can be used for a LO. Several types of fibers  can be used such as few mode fibers (FMF), multicore fibers (MCF) and even a bundle of single-mode fibers (SMF). Multimode optical fibers are used to implement high-dimensional QKD \cite{Canas2017,Ding2017,Bal19}.

%%%%%%%%%%%%%%%%%%%%%%%%%%%%%%

%\subsection{N-dimensional quantum states and protocol}
The protocol is as follows: for each single event, Alice randomly chooses the same quadrature (or {\em{basis}}) $\mathcal{E}$ or $\mathcal{P}$ for all modes. If she chooses the first quadrature $\mathcal{E}$, she applies a random phase $\phi_k={0,\pi}$ to each mode. On the contrary, she applies the random phases
$\{ \pi/2, 3\pi/2 \}$ to select the second quadrature $\mathcal{P}$. In short,  Alice can send the following states 
\begin{equation}
\vert \Psi\rangle_{X}=\vert \alpha_{1} ... \alpha_{N}\rangle_{X}= {\ket{(-1)^{\nu_{1}}a_{1}..(-1)^{\nu_{N}}a_{N}}}
\end{equation}
\begin{equation}
\vert \Psi\rangle_{P}=\vert \alpha_{1} ... \alpha_{N}\rangle_{P}= {\ket{(-1)^{\nu_{1}}ia_{1}..(-1)^{\nu_{N}}ia_{N}}},
\end{equation}
   %of the first quadrature can never combine phases with integer multiples of $\pi$ with half integer ones (f.e. in 4 dimensions the state $\ket{\alpha, i\alpha}$ is not allowed). 
   with $\nu_{k}=\{0,1\}$. For example, if $N=1$ we will have the two states $\vert \pm a_{1}\rangle$,  in the $\mathcal{E}$-quadrature, and   $\vert \pm ia_{1}\rangle$ in the $\mathcal{P}$-quadrature. These four states were used in the seminal work by Namiki and Hirano \cite{Namiki2003} for   $N$$=$$1$; however, for $N=2$ we have,   in the particular case of $a_{1}=a_{2}$ the following eight states:  $\vert \pm a_{1}\pm a_{1}\rangle$ and $\vert \pm ia_{1}\pm ia_{1}\rangle$, that is, we have $2^{2(=N)+1}$$=$$8$ states. Afterwards,  Bob randomly chooses to measure the first or the second quadrature by a  homodyne detection, that is, one of the  conjugated basis $\mathcal{E}/\mathcal{P}$ is selected  by applying the same phase $\phi_{B}=0,\frac{\pi}{2}$ to the local oscillator for every mode. If Bob  performs an homodyne measurement for all modes with  phase $\phi_{B}=0$ a  product of weak coherent states  in the optical field $\hat{\mathcal{E}}$ basis for each mode will be measured, otherwise  if $\phi_{B}=\frac{\pi}{2}$ a measurement in the optical field momentum $\hat{\mathcal{P}}$ basis is made. Let us recall  that these optical field operators are related to the absorption operator as follows $\hat{a}_{k} = \hat{\mathcal{E}}_{k} + i\hat{\mathcal{P}}_{k} $, therefore $[\hat{\mathcal{E}}_{k},\hat{\mathcal{P}}_{k^{'}}] = \frac{i}{2} \delta_{kk^{'}} $ .

\begin{figure}
\centering
\includegraphics[width=0.7\textwidth]{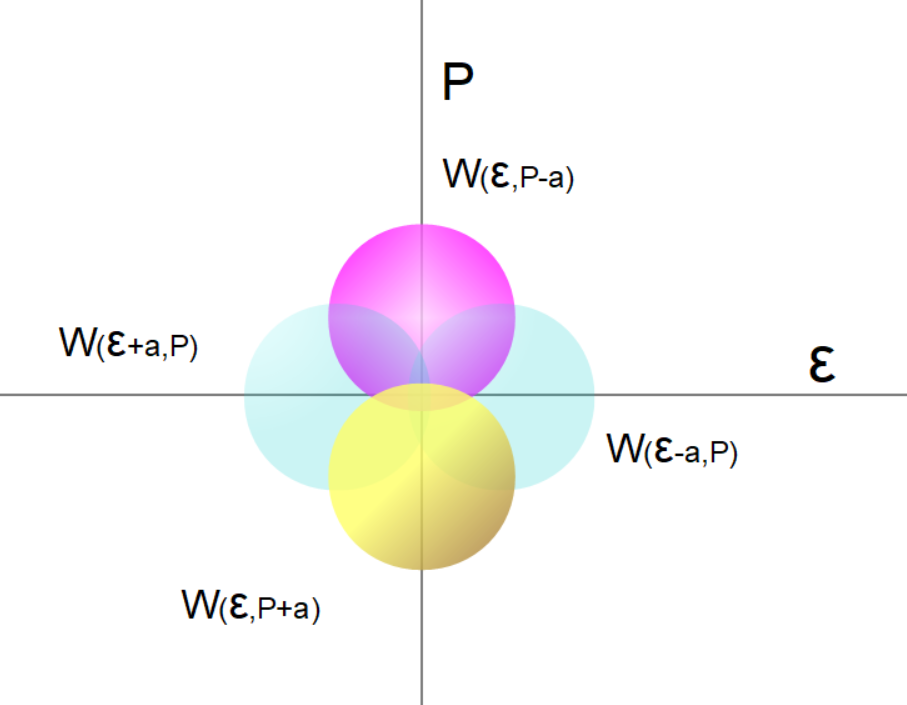}
\caption{The four weak coherente states for  $N$$=$$1$ mode represented with the Wigner function in the phase space $\mathcal{E}/\mathcal{P}$. }
\label{fig1}
\end{figure}

At this point, it is interesting to note that coherent states form a overcomplete set of states, and therefore a set of mutually unbiased bases (MUBs) can not be exactly defined  as in DV-QKD. Nevertheless, it can be seen that in the phase space $\mathcal{E}/\mathcal{P}$ these states  remind us of MUBs although they overlap with each other; indeed, in Fig.1, it is shown for   $N$$=$$2$ that there is an identical overlap  between states in $\mathcal{E}$ basis and states in $\mathcal{P}$ basis as required with  unbiased bases.  %smaller between the states in the $\mathcal{E}$ basis than for states in $\mathcal{E}$ and $\mathcal{P}$ basis. In addition, these overlaps will conform one of the main advantages of the protocol for the security as will be shown in the next sections.

\begin{figure}
\centering
\includegraphics[width=0.7\textwidth]{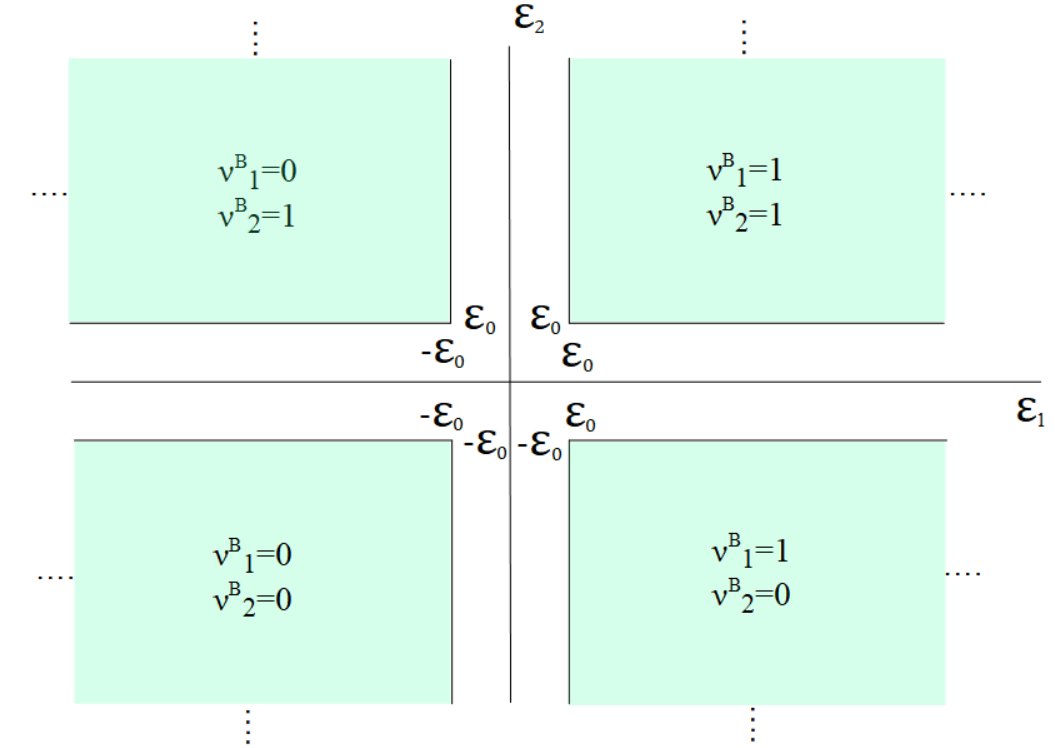}
\caption{The values $\nu_{1}^{B}$ and $\nu_{2}^{B}$ for $N$$=$$2$ are indicated,  within the green regions (threshold regions) with squared frontiers in the two-dimensional first quadrature space $\mathcal{E}_{1}\mathcal{E}_{2}$,  according to the values of the $(\alpha_{1},\alpha_{2})$ of a coherent state $\vert \alpha_{1}\alpha_{2}\rangle$. We set the thresholds $\mathcal{E}_{o,1}=\mathcal{E}_{o,2}=\mathcal{E}_{o}$. }
\label{fig0}
\end{figure}

Next, Alice and Bob retain the pulses with coincident basis obtaining a sifted key. Due to overlapping among these states, a set of thresholds $\{\mathcal{E}_{o,1},...,\mathcal{E}_{o,N}\}$ (for example in $\mathcal{E}$ basis) have to be selected by Bob to discriminate between states and therefore a bit assignment criteria $\nu^{B}_{k}$ $(k=1,...,N)$ is constructed, that is,   

\begin{equation}
\label{criteria_bob}
\nu^{B}_{k}=
    \begin{cases}
        1 & \text{if} \hspace{1.0mm} \mathcal{E}_{k}\geq \mathcal{E}_{o,k} \hspace{1.0mm} \text{and } \hspace{1.0mm}  |\mathcal{E}_{l}|> \mathcal{E}_{o,l}  \hspace{1.0mm}  \forall l \neq k  \\
        0 & \text{if } \hspace{0.0mm} \mathcal{E}_{k}< -\mathcal{E}_{o,k}  \hspace{1.0mm} \text{and } \hspace{1.0mm}  |\mathcal{E}_{l}|> \mathcal{E}_{o,l}  \hspace{1.0mm}  \forall l \neq k \\
        \text{none} & \text{if} \quad \exists \, l \hspace{0.95mm} /\hspace{0.5mm} \vert\mathcal{E}_{l}\vert< \mathcal{E}_{o,l} \hspace{1.0mm}  
        
    \end{cases}
\end{equation}
where $\mathcal{E}_{k}$ is the result of Bob's measurement. If any optical field value does not fulfill the criteria then no bits are distilled. These thresholds can be selected arbitrarily, but for a practical case squared frontiers are selected, as shown  in Fig.\ref{fig0}, and accordingly thresholds regions are selected.

On the other hand, when Alice announces the basis, that is, the quadrature used,  then the  $2^{N+1}$-dimensional density operator is reduced for each of the basis $\mathcal{E}/\mathcal{P}$  to 
\begin{equation}
\label{rho1}
    \hat{\rho}_{1}=\frac{1}{2^{N}}\sum_{\nu_{1}..\nu_{N}=0,1}{\ket{(-1)^{\nu_{1}}a_{1}..(-1)^{\nu_{N}}a_{N}}}  \bra{(-1)^{\nu_{1}}a_{1}..(-1)^{\nu_{N}}a_{N}}
\end{equation}
\begin{equation}
    \hat{\rho}_{2}=\frac{1}{2^{N}}\sum_{\nu_{1}..\nu_{N}=0,1}{\ket{(-1)^{\nu_{1}}ia_{1}..(-1)^{\nu_{N}}ia_{N}}} \bra{(-1)^{\nu_{1}}ia_{1}..(-1)^{\nu_{N}}ia_{N}},
    \label{rho2}
\end{equation}
therefore Bob's task will be to distinguish as best as possible between the $2^{N}$ states of the first ($\mathcal{E}$) or the second ($\mathcal{P}$) quadrature. The criteria given by Eq.(\ref{criteria_bob}) can be used to perform the mentioned task by using a balanced homodyne detection.

For the sake of obtaining a quantitative description of the above task,  we write the general $N$-dimensional wave function of an arbitrary coherent state in the optical field representation, together with its  probability density,   and  under an apropriated optical-quantum unities \cite{schleich01, Loudon1983}, that is, 
\begin{equation}
\label{N-wavefunction}
     \psi(\boldsymbol{\mathcal{E}})=\left( \dfrac{2}{\pi}\right)^{N/4}\prod_{k=1}^{N}{\exp[{-(\mathcal{E}_{k}-\mathcal{R}(\alpha_{k}))^{2}}+2i\mathcal{I}(\alpha_{k})(\mathcal{E}_{k}-\mathcal{R}(\alpha_{k}))] }
\end{equation}
\begin{equation}
\label{N-prob}
    |\psi(\boldsymbol{\mathcal{E}})|^2 = \left( \dfrac{2}{\pi}\right)^{N/2}\prod_{k=1}^{N}{\exp[{-2(\mathcal{E}_{k}-\mathcal{R}(\alpha_{k}))^{2}}]},
\end{equation}
with ${\boldsymbol{\mathcal{E}}}=(\mathcal{E}_{1},..,\mathcal{E}_{N})$. The real and imaginary part of $\alpha_{k}$ are respectively $\mathcal{R}(\alpha_{k})=a_{k} \cos\phi_{k}$ and $\mathcal{I}(\alpha_{k})=a_{k}\sin\phi_{k}$. They represent an $N$-dimensional Gaussian probability distribution. Note that in this $\mathcal{E}$-representation the following equivalence is obtained $\mathcal{R}e(\alpha_{k})=\mathcal{E}_{k}$, that is, the mean optical field  \cite{schleich01, Loudon1983}.

\begin{figure}[h]
\centering
     \begin{subfigure}[h]{0.47\textwidth}
         
         \includegraphics[width=1\textwidth]{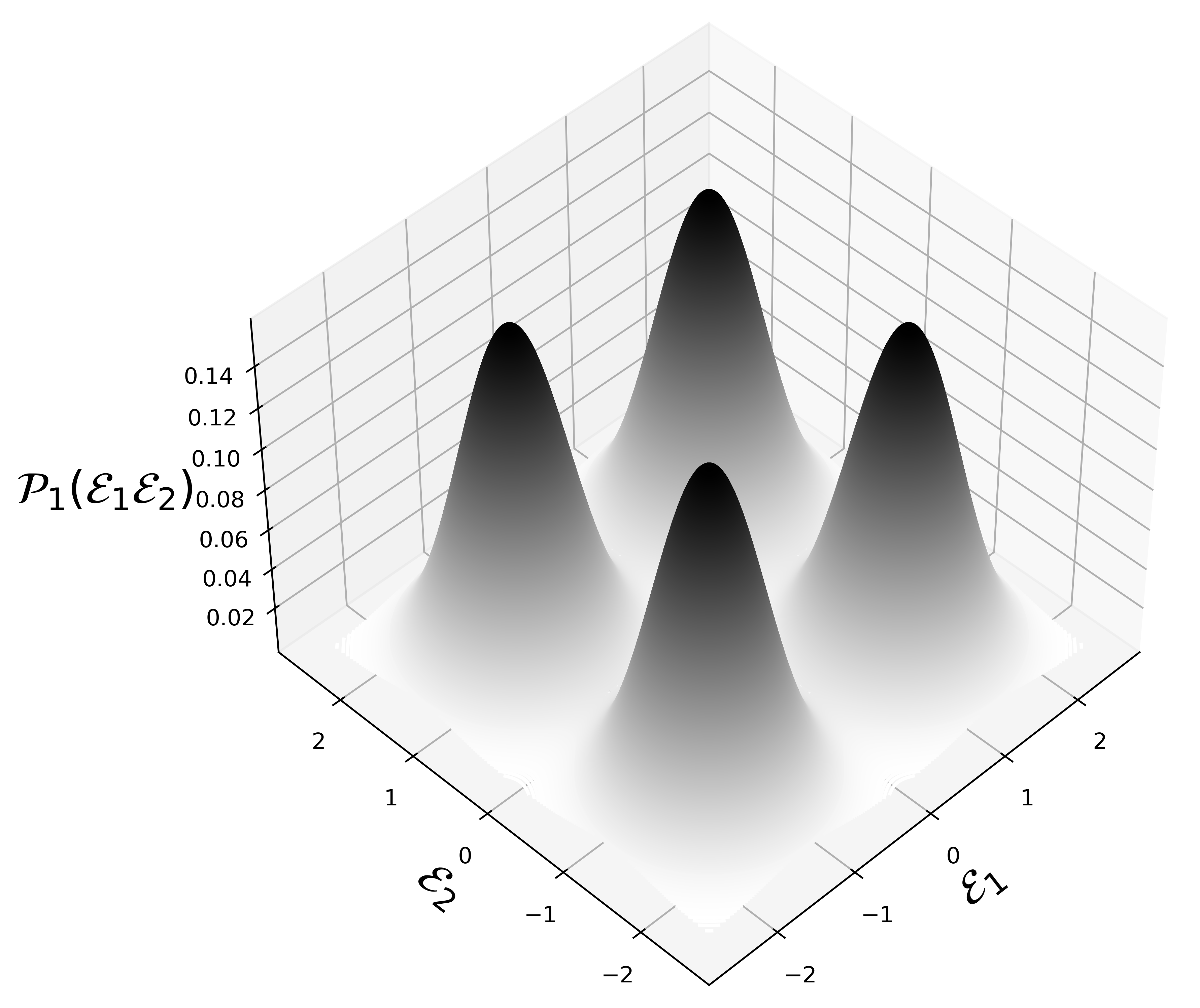}
         \caption{}
         \label{fig2a}
         
     \end{subfigure}
     \hfill
     \begin{subfigure}[h]{0.47\textwidth}
         
         \includegraphics[width=1\textwidth]{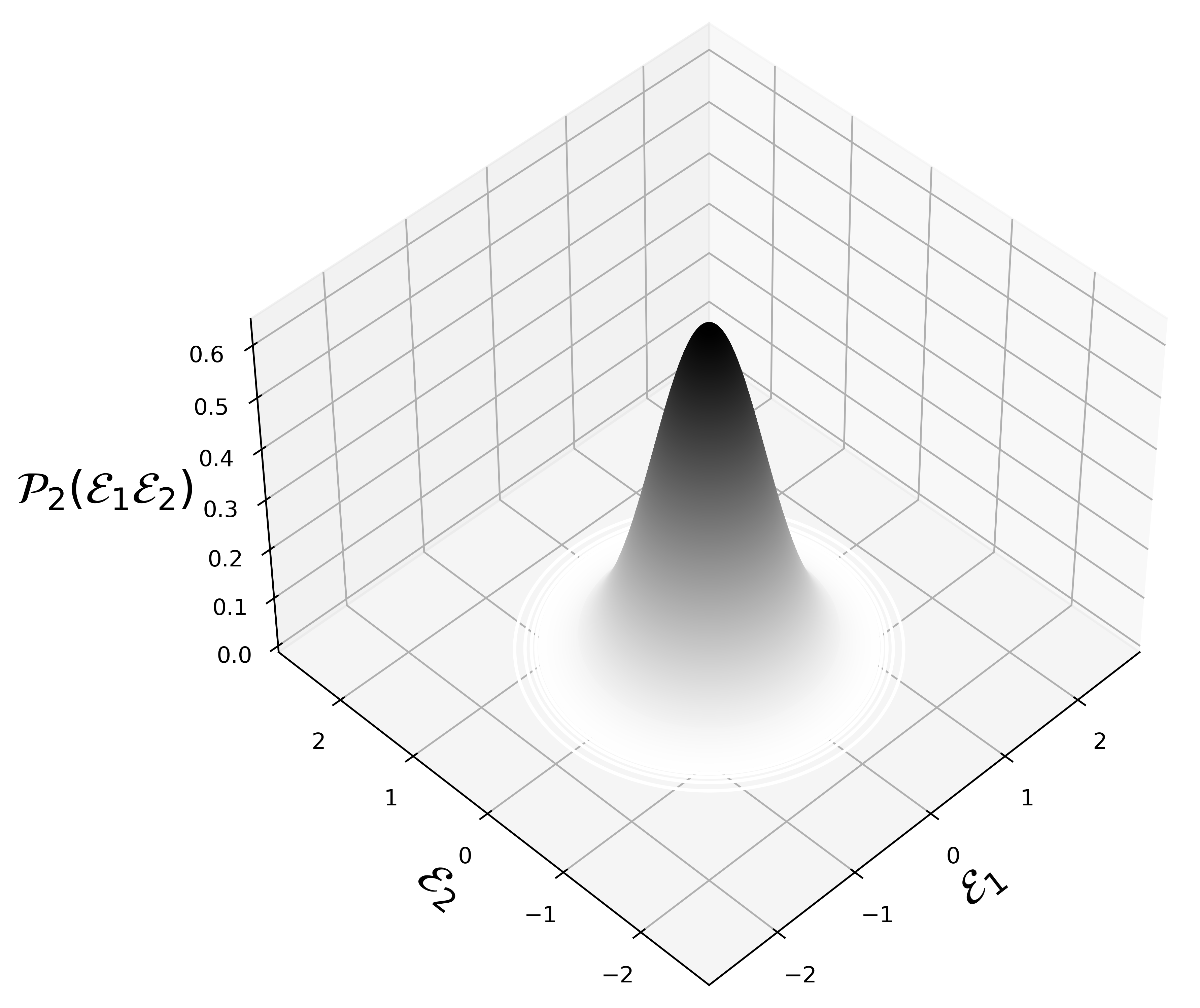}
         \caption{}
         \label{fig2b}
         
     \end{subfigure}
\caption{Representation of the probability distribution of the two bases $\boldsymbol{\mathcal{E}}/\boldsymbol{\mathcal{P}}$ in $\mathcal{E}_{1}\mathcal{E}_{2}$ first quadratures : (a)   $\mathcal{P}_{1}(\mathcal{E}_{1},\mathcal{E}_{2})$  for  the set of states $\{$$\ket{a_1 a_2}$, $\ket{-a_1 a_2}$, $\ket{-a_1 -a_2}$, $\ket{a_1 -a_2}$$\}$  and (b) the centered bidimensional Gaussian  $\mathcal{P}_{2}(\mathcal{E}_{1},\mathcal{E}_{2})$ represents the four $\boldsymbol{\mathcal{P}}$ states $\{$$\ket{i a_1 i a_2}$, $\ket{-i a_1 i a_2}$, $\ket{-i a_1 -i a_2}$, $\ket{i a_1 -i a_2}$$\}$ .}
\label{fig2}
\end{figure}

 %From now on and without loss of generality, we will suppose that Alice sends states in the $\mathcal{E}$ basis. 
Therefore, if Alice uses the phases $\phi_{k}=0,\pi$  Gaussian distributions centered in $\pm a_{k}$ are obtained in the first quadrature, and if Alice uses the values $\phi_{k}=\frac{\pi}{2},\frac{3\pi}{2}$ then Gaussian distributions centered in $\pm a_{k}$ are obtained in the second quadrature,  however, in this case  it is obtained indistinguishable Gaussian distributions centered around $a_{k}=0$ in the first quadrature. Therefore the $N$-dimensional probability distribution of the optical field $\mathcal{E}$ measured by Bob is written as follows in the $\mathcal{E}$ basis
\begin{equation}
\label{projection_rho1}
    \braket{{\boldsymbol{\mathcal{E}}}|\hat{\rho_{1}}|{\boldsymbol{\mathcal{E}}}}= \frac{1}{2^{N}}\sum_{\nu_{1}..\nu_{N}=0,1}\vert \psi(\mathcal{E}_{1},(-1)^{\nu_{1}}a_{1})  ...\psi(\mathcal{E}_{N},(-1)^{\nu_{N}}a_{N})\vert^{2}
\end{equation}
\begin{equation}
\label{projection_rho2}
   \braket{{\boldsymbol{\mathcal{E}}}|\hat{\rho_{2}}|{\boldsymbol{\mathcal{E}}}}= \vert \psi(\mathcal{E}_{1},0)  ...\psi(\mathcal{E}_{N},0)\vert^{2}. 
\end{equation}
where we have used $\mathcal{R}(\alpha_{k})=(-1)^{\nu_{k}}a_{k}$. 
%employed the next redefinition for scalar variables:
%\begin{equation*}
%\psi(\mathcal{E},a) =\left(\dfrac{2}{\pi}\right)^{1/4} \exp[{-(\mathcal{E}-a))}^{2}]
%\end{equation*}
Next, to illustrate these general wavefunctions,  a graphical representation of their quadrature probabilities is shown in Fig.\,\ref{fig2} for the case $N=2$.  Moreover, their expressions  in the optical field space $\mathcal{E}_{1}\mathcal{E}_{2}$ are given by
\begin{equation*}
\braket{\mathcal{E}_{1}\mathcal{E}_{2}|\hat{\rho_{1}}|\mathcal{E}_{1}\mathcal{E}_{2}}\equiv \mathcal{P}_{1}(\mathcal{E}_{1},\mathcal{E}_{2})=  \frac{1}{2\pi} \ [ \exp{[-2(\mathcal{E}_{1}-a_{1})^{2}-2(\mathcal{E}_{2}-a_{2})^{2}]}+
\end{equation*}
\begin{equation*}
\exp{[-2(\mathcal{E}_{1}-a_{1})^{2}-2(\mathcal{E}_{2}+a_{2})^{2}]}+\exp{[-2(\mathcal{E}_{1}+a_{1})^{2}-2(\mathcal{E}_{2}-a_{2})^{2}]}+
\end{equation*}
\begin{equation}
+\exp{[-2(\mathcal{E}_{1}+a_{1})^{2}-2(\mathcal{E}_{2}+a_{2})^{2}]}\}
\end{equation}
\begin{equation}
  \hspace{0cm}  \braket{\mathcal{E}_{1}\mathcal{E}_{2}|\hat{\rho_{2}}|\mathcal{E}_{1}\mathcal{E}_{2}}\equiv \mathcal{P}_{2}(\mathcal{E}_{1},\mathcal{E}_{2})= \frac{2}{\pi} \exp{[-2(\mathcal{E}_{1}^{2}+\mathcal{E}_{2}^{2})]}.
\end{equation}

We must stress that the bit errors will come from the overlaps  between  the weak coherent states, which are distributed around $\boldsymbol{\mathcal{E}}=\boldsymbol{0}$. However, a relevant set of such errors can be avoided by using a proper threshold such as given by Eq.\,(\ref{criteria_bob}).

%%%%%%%%%%%%%%%%%%%%%%%%%%

\section{$2^{N}$ dimensional postselection efficiency and Bit Error Rate}
\begin{figure*}[h]
     \centering
     \begin{subfigure}{0.45\textwidth}
         
         \includegraphics[width=1\textwidth]{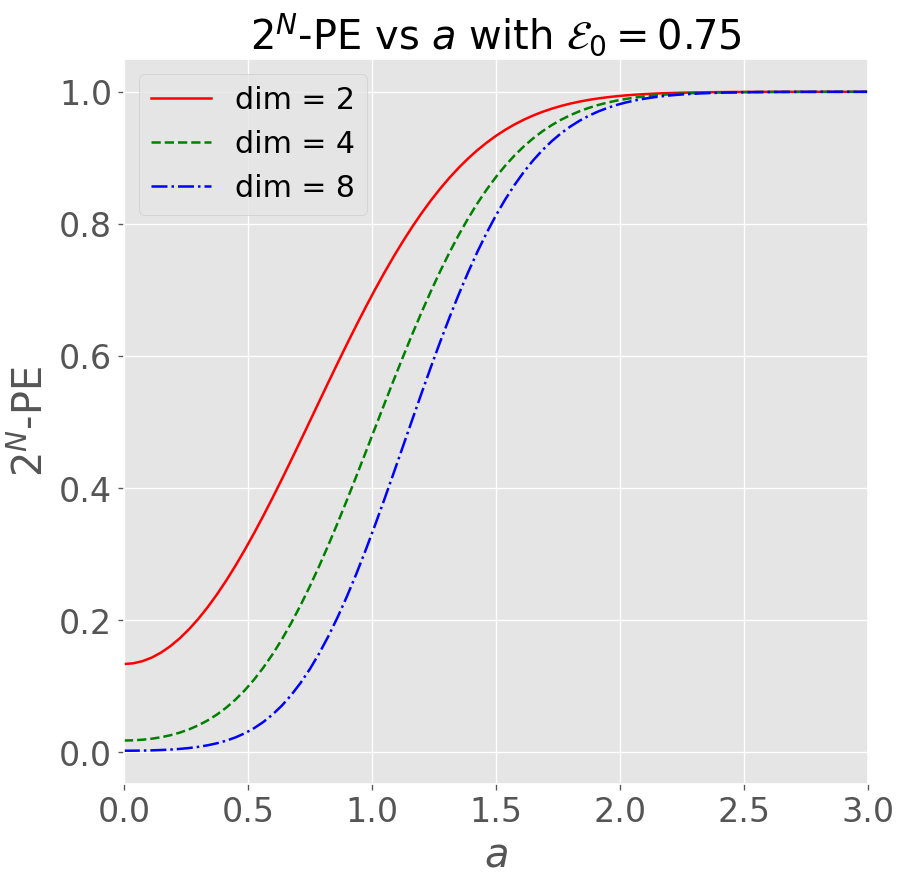}
         \caption{}
         \label{fig3a}
         
     \end{subfigure}
     \hfill
     \begin{subfigure}{0.45\textwidth}
         
         \includegraphics[width=1.0\textwidth]{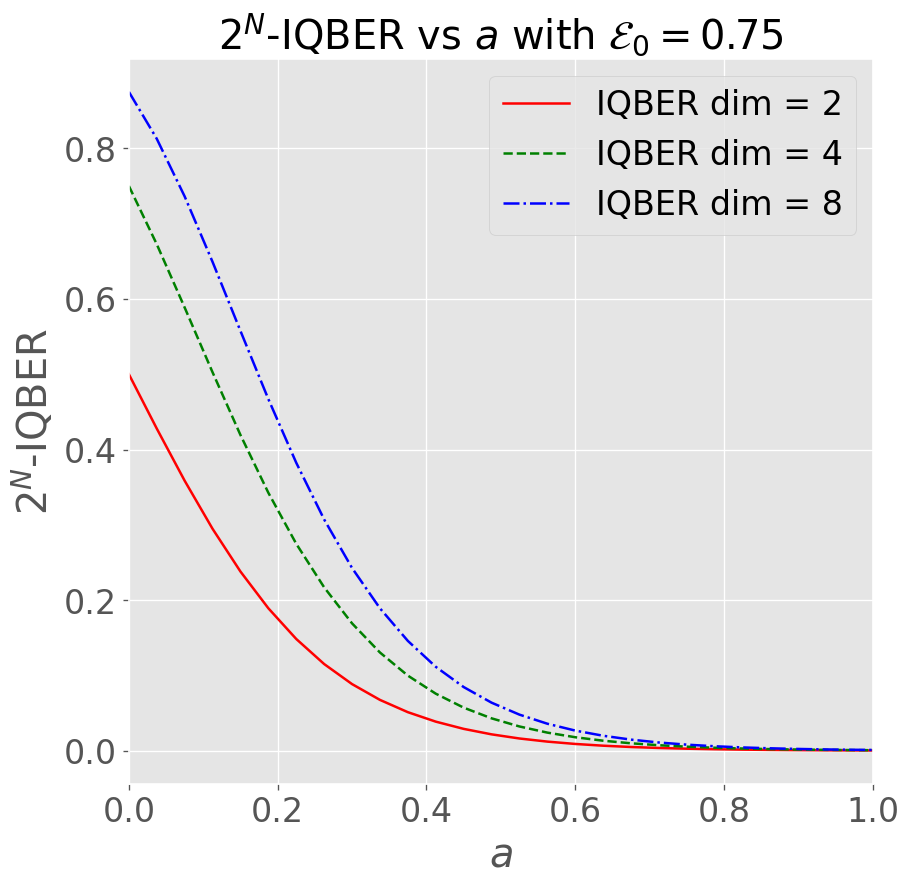}
         \caption{}
         \label{fig3b}
     \end{subfigure}
     
        \caption{ (a) Representation of $2^{N}$-PE for $N=1, 2, 3$ modes as a function of the mean optical field in each mode, that is, $a_{1}$\,$=$\,$a_{2}$\,$=$\,$a_{3}$\,$=$\,$a$ for a  typical threshold value  $\mathcal{E}_{o,1}$\,$=$\,$\mathcal{E}_{o,2}$\,$=$\,$\mathcal{E}_{o,3}$\,$=$\,$ \mathcal{E}_{o} $\,$=$\,$ 0.75$. (b) Representation of IQBER for $N=$ 1, 2, and 3 modes and  $\mathcal{E}_{0} = 0.75$; the horizontal axis again represents the mean optical field $a$ in each mode.}
        \label{fig3}
\end{figure*}

In this section we extend the results for the two-dimensional case ($N$$=$$1$ mode) obtained by Namiki and Hirano   \cite{Namiki2003} to the $2^{N}$ dimensional case ($N$ modes). Thus, the probability that Alice and Bob share a  bit value according to the criteria given by Eq.(\ref{criteria_bob}) is defined as the   $2^{N}$-Postselection Efficiency ($2^{N}$-PE) whose expression can be written as follows in the $\mathcal{E}$ basis 

\begin{equation}
P(\boldsymbol{\mathcal{E}}_{0},\boldsymbol{a}) = \int_{\textit{D}}^{}{\braket{\boldsymbol{\mathcal{E}}|\hat{\rho_{1}}|\boldsymbol{\mathcal{E}}}}
\end{equation}
where $\textit{D}$ is the domain of integration formed by all the threshold regions, which,  in our case and from a numerical point of view, will be the hypervolume of the difference between a $N$ dimensional hypercube and a $N$ dimensional cross delimited by the thresholds $\boldsymbol{\mathcal{E}}_{0}$ (see, for example,  Fig.\ref{fig0} for $N$$=$$2$). By using this squared frontiers as thresholds and the symmetry properties of the Gaussian integrals, we obtain, after a long but straightforward calculation,  the following expression
\begin{equation}
\label{PE-N}
    P(\boldsymbol{\mathcal{E}}_{0},\boldsymbol{a}) = \prod_{k=1}^{N}{P_{k}(\mathcal{E}_{0,k},a_{k})}, 
\end{equation}
with  ${\boldsymbol{\mathcal{E}_{0}}} =  (\mathcal{E}_{0,1},..,\mathcal{E}_{0,N})$, $\boldsymbol{a}=((-1)^{\nu_{1}}a_{1},..,(-1)^{\nu_{N}}a_{N})$ and $P_{k}(\mathcal{E}_{0,k},\alpha_{k})$ the one dimensional postselection efficiency given  by
\begin{equation}
 P_{k}(\mathcal{E}_{0,k},a_{k})=\frac{1}{2}[\erfc{(\sqrt{2}(\mathcal{E}_{0,k} - a_{k}))}+\erfc{(\sqrt{2}(\mathcal{E}_{0,k} + a_{k}))}].
\label{PE-1} 
\end{equation}
Note that  for $\mathcal{E}_{0,k}=0$ every correct-basis pulse gives a bit value, therefore  $P({\boldsymbol{\mathcal{E}_{0}}},\boldsymbol{a})=1$.  In Fig.\ref{fig3a} plots of $2^{N}$-PE are shown  by  using typical threshold values and for  several dimensions.  %\textcolor{red}{Analogously ??? (non comparable) to a multichannel protocol, the total postefficiency is the product of   each individual channel postefficiency}.

Finally, there will be an intrinsic quantum bit error rate ($2^{N}$-IQBER) due to overlapping, that is, the probability that a measurement that Bob considered valid (that is, postselected), was
actually wrong, as  for example, in the $\mathcal{E}$ basis,  when Bob measures an output $\mathcal{E}<-\mathcal{E}_{0,k}$ for one or several $k$,  when Alice sends the state $\vert a_{1} \cdot \cdot \cdot  a_{N}\rangle$. Such probability can be calculated from the complemetary one: the probability that a postselected measurement was right, which is ratio between the probability of $\mathcal{E}$ being within the $2^{N}$ threshold regions chosen by Alice, according to the quantum state selected,  and the probability of being in any point of the domain D defined above (the latter is, in fact, $2^{N}$-PE), that is,
\begin{equation}
   \hspace{-0,31cm} q({\boldsymbol{\mathcal{E}_{0}}},\boldsymbol{a}) =  1-\prod_{k=1}^{N}{\frac{\erfc{(\sqrt{2}(\mathcal{E}_{0,k} - a_{k}))}}{2P_{k}(\mathcal{E}_{0,k},a_{k})}} = 1-{\frac{ \prod_{k=1}^{N}\erfc{(\sqrt{2}(\mathcal{E}_{0,k} - a_{k}))}}{2^{N}P({\boldsymbol{\mathcal{E}_{0}}},\boldsymbol{a})}}
\label{iqber}
\end{equation}
where the second term on the rightside %$\frac{1}{2^{N}} \prod_{k=1}^{N}{\frac{\erfc{(\sqrt{2}(\mathcal{E}_{0,k} - a_{k}))}}{P_{k}(\mathcal{E}_{0,k},a_{k})}}$
 represents the probability of obtaining a correct measurement in the $2^{N}$ thresholds regions.  
Next, we write, from  Eq.(\ref{iqber}) and as illustrative examples,  the expression of the IQBERs for  $N$$=$$1$ and $N$$=$$2$ cases  with  the same thresholds and mean photon number in each mode, that is,
\begin{equation}
q(\mathcal{E}_{0},a) = 1- \frac{1}{2} \frac{\erfc{(\sqrt{2}(\mathcal{E}_{0} - a))}}{P(\mathcal{E}_{0},a)} = \frac{1}{2P(\mathcal{E}_{0},a)} \erfc{(\sqrt{2}(\mathcal{E}_{0} + a))}
\end{equation}
\begin{equation*}
q(\mathcal{E}_{0},\mathcal{E}_{0},a,a) = 1- \frac{1}{4} \frac{\erfc^{2}{(\sqrt{2}(\mathcal{E}_{0} - a))}}{P^{2}(\mathcal{E}_{0},a)} =
\end{equation*}
\begin{equation}
\frac{1}{4P^{2}(\mathcal{E}_{0},a)} [2\erfc{(\sqrt{2}(\mathcal{E}_{0} - a))}\erfc{(\sqrt{2}(\mathcal{E}_{0} + a))}+\erfc^{2}{(\sqrt{2}(\mathcal{E}_{0} + a))}]
\end{equation}
where the Eq.(\ref{PE-1}) is used in both cases.   Finally,
 %We must stress that this multidimensional IQBER differs from a multichannel IQBER which would be given by  the sum $\sum_{k=1}^{N}{[1-\frac{1}{2}\frac{\erfc{(\sqrt{2}(\mathcal{E}_{0,k} - \alpha_{k}))}}{P_{k}(\mathcal{E}_{0,k},a_{k})}]}$. 
in Fig.\ref{fig3b} is shown a comparison of $2^{N}$-IQBER for $N=$1, 2 and 3 modes and with a typical threshold   $\mathcal{E}_{0} = 0.75$. Likewise, note that for  low energy limit $\sqrt{n_{i}}=a_{i} \ll1$, the $2^{N}$-IQBER tends to $\frac{2^{N}-1}{2^{N}}$ for arbitrary thresholds. That is, the measure becomes fully random when the state  approaches to vacuum, as expected.

%%%%%%%%%%%%%%%%%%%%%%%%%%%%%%%%%%%%%%%%%%%%%%
%%%%%%%%%%%%%%%%%%%%%%%%%%%%%%%%%%%%%%%%%%%%%

\section{Autocompensating HD-DM-CV-QKD system}

The protocol described above will be implemented by a  scheme assisted by polarization, that is, a multimode optical fiber will be used but quantum states (signal) and  LO  will be excited in different polarization modes.  As commented, several types of fibers  can be used such as few mode fibers (FMF), multicore fibers (MCF) and even a bundle of single-mode fibers (SMF).  In all these cases, particularly in SMF,  it can be  assumed  that there is no  spatial coupling, however, polarization  modal coupling limits their applicability for long distance links. Such coupling is due to slow external random perturbations (mechanical, thermical...)  or to waveguide birrefrigence (imperfections) of the optical fibers. 

\begin{figure*}[h]
\centering
\includegraphics[width=1\textwidth]{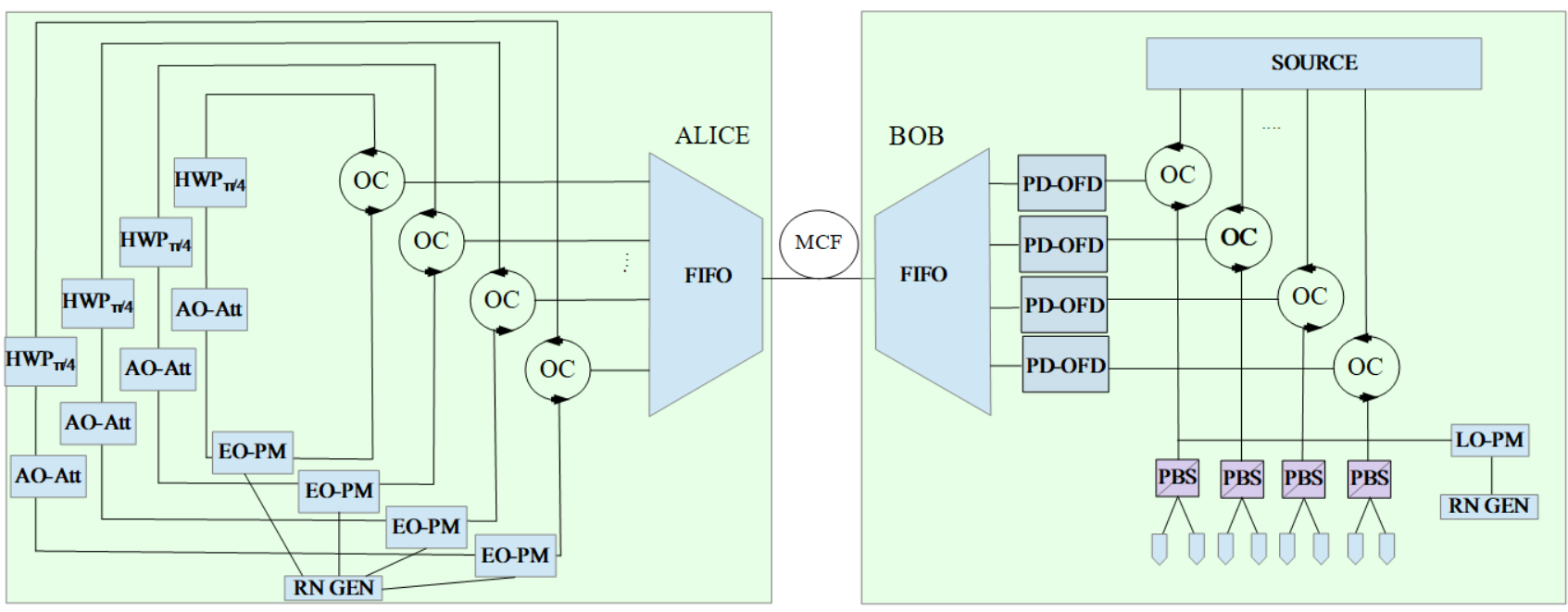}
\caption{Autocompensating  system for the HD-DM-CVC-QKD protocol. List of elements: MCF (multicore optical fiber). ALICE: FIFO (Fan-in/Fan-out Coupler), OC (optical circulator), HWP (half wave plate),  AO-AT (acusto-optical attenuator), RN-GEN (random number generator), EO-PM (electrooptical phase modulator). BOB: Source (laser), OC (optical circulator), PD-OFD (polarization dependent optical fiber delay),  FIFO (Fan-in/Fan-out Coupler), RN-GEN (random number generator), LO-PM (local oscillator phase modulator), PBS (polarization beam splitter).}
\label{fig4}
\end{figure*}

Although an active compensating device could be used, a  passive compensation (autocompensation) QKD supposes a very practical solution to overcome these caveats. It consists of  propagating the light  along  a round trip, and  at the point where the light returns a proper linear transformation is introduced to achieve autocompensation on the return trip.  Several systems have been proposed by using Faraday mirrors  \cite{Kawamoto2005}, Half-Wave-Plates (HWP)  for bulk devices \cite{Bal19},  directional couplers for integrated devices  \cite{Bal19,Linares2021} and  non-linear transformations \cite{PRA2021}. In this case HWPs are a satisfactory solution  for a passive autocompensating of the HD-DM-CV-QKD protocol as shown in Fig.\ref{fig4}.

\begin{figure}[h]
     \centering
     \begin{subfigure}{0.45\textwidth}
         
         \includegraphics[width=0.9\textwidth]{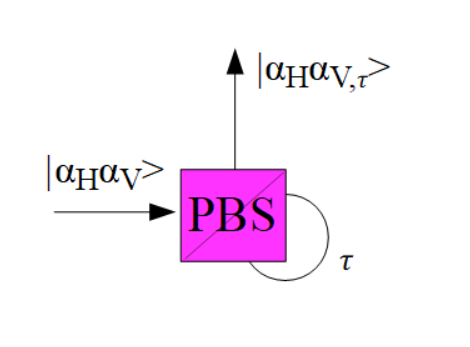}
         \caption{}
         \label{fig5a}
     \end{subfigure}
     \hfill
     \begin{subfigure}{0.45\textwidth}
         \includegraphics[width=0.9\textwidth]{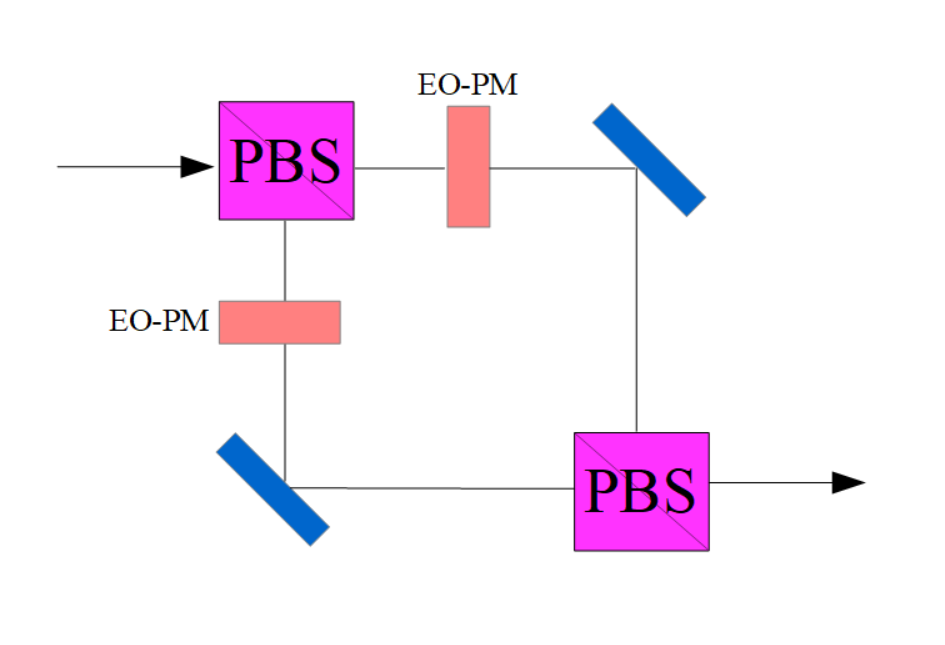}
         \caption{}
         \label{fig5b}
     \end{subfigure}
        \caption{(a) Sketch of the optical fiber delay (PD-OFD): it splits an incoming optical signal into an horizontal linearly polarized coherent state (signal) and a  $\tau$-delayed vertical linearly polarized coherent state (local oscillator). (b) Sketch of electrooptical phase modulator where polarization components must be split and recombined, since each EO-PM acts on only one component, although they both do so synchronously. This global phase is only applied to the signal, turning off the EO-PMs afterwards for a non modulation of the incoming LO. }
        \label{fig5}
\end{figure}

The autocompensating process is as follows: initially,  Bob emits a multimode coherent pulse  with the same mean number of photons in each mode. It can be obtained by using a coherent source (a laser) and can be coupled to $N$ SMF optical fibers. The multimode  state is sent through  optical circulators (OC) that derive it to several  polarization dependent optical fiber delayers (PD-OFD) as shown in Fig.\ref{fig5a}. Here the signal separates into two linear polarized pulses: the horizontal one is called signal, the vertical local oscillator (LO); this last one is delayed a time $\tau$. Thus, Alice will be able to modify the signal pulse to introduce the key without altering the LO, even though she will receive their polarizations disturbed. Next, a Fan-in/Fan-out (FIFO) coupler  allows the $N$ signals to propagate from single mode fibers (SMF) into one multicore fiber (MCF). In short, the output state from Bob can be written as follows
\begin{equation}
\vert \Psi_{o}\rangle= \vert \alpha_{o1H} ... \alpha_{oNH}\rangle \vert \alpha_{o1V} ... \alpha_{oNV}\rangle_{\tau}
\end{equation}
where we have assumed that $\alpha_{oiH}=A\exp({i\phi_{oi}}), \alpha_{oiV}=A\exp({i\phi_{oi}})$, that is, the coherent states in each polarization space have  the same phase $\phi_{oi}$. 

We must stress that under propagation each coherent state is perturbated, for example, $\vert \alpha_{o1H}\rangle\rightarrow  \vert \alpha'_{o1H} \alpha'_{o1V}\rangle$. When the signal arrives at Alice, after a FIFO and the corresponding OCs, each perturbated polarization coherent state of the product state passes through a  Half Wave Plate with angle $\pi/4$ (HWP$_{\pi/4}$)  and therefore polarization modes are permuted (H$\rightarrow $V, V$\rightarrow$H) (that is, a logic X gate is applied). This will allow polarization perturbations to be effectively compensated when the pulses reach Bob  \cite{Bal19,PRA2021}.
 In order to attenuate the signal, an  acusto-optical optical attenuator (AO-Att) is employed by  making use of the delay between polarization modes, therefore $A\rightarrow a$, with $a\ll A$. After AO-Att the signal gets weakened and therefore entering the quantum regime, that is, a multimode weak coherent states is achieved. After the signal, the LO passes through and there should be enough time for the AO-Att to switch off, not affecting the LO which keeps on staying in classical regime. Similarly,  Alice performs a phase modulation by using a fast electrooptical phase modulator (EO-PM), which applies a phase $\phi_{A}$ to the signal (Fig.6b). After this, as commented,  the  incoming LO does not  get modulated by the  EO-PM because  such modulator quicky turns off after the signal is modulated. 
Along  the path between Alice and Bob, both signal and LO pulses remain delayed a time $\tau$, but their polarization changes. When the pulses reach the PD-OFD of Bob's station, the signal polarization becomes vertical and that of the LO becomes horizontal. Consequently, in this case, the delay is applied onto the non primarily delayed pulse. Therefore, autocompensation is succesfully performed. The state reaching the detection system at Bob is given by
\begin{equation}
\vert \Psi\rangle=\vert \alpha_{1V}    ...  \alpha_{NV}  \rangle\, \vert \alpha_{1H}   ...  \alpha_{NH} \rangle
\end{equation}
with 
\begin{equation}
\alpha_{iV}=(-1)^{\nu_{i}}a\,\exp({i\phi_{i}}),  \alpha_{iH}=A\,\exp({i\phi_{i}}) \hspace{2mm} \mathcal{E} \text{ basis}
\end{equation}
or 
\begin{equation}
\alpha_{iV}=(-1)^{\nu_{i}}a\,\exp({i(\phi_{i}+\frac
{\pi}{2})}),  \alpha_{iH}=A\,\exp({i\phi_{i}}) \hspace{2mm} \mathcal{P} \text{ basis}
\label{pi2}
\end {equation}

The final measuring step begins with a PBS at $45^{o}$ which splits again the light into signal and LO, both polarized at the contrary as Bob's output, due to the permutation of the HWP. Because of this, several EO-PM are placed where the horizontal polarization component comes out to apply a random phase $\phi_{B}$ to the LO (the same for all modes). Finally, a homodyne detection with a a pair of photodetectors for each mode, which are capable of measuring the homodyne current. After all of this measuring process a secure key can be distilled. % making use of the two usual addicional steps to remove erroneous bits known as information reconciliation and privacy amplification. 
%As can be seen in (\ref{pi2}), if Alice chooses the $\mathcal{P}$ basis; an additional $\frac{\pi}{2}$ phase must be included in $\alpha_{iV}$.

%%%%%%%%%%%%%%%%%%%%%%%%%%%%%%%%%%%%%%%%%%%%%%
%%%%%%%%%%%%%%%%%%%%%%%%%%%%%%%%%%%%%%%%%%%%%

\section{Simultaneous homodyne measurement attack}

The primary aim of this section is to analyze the security of the HD-CV-QKD protocol. For that, we analyze, for example,  an intercept and resend attack implemented by a simultaneous  homodyne measurement that is, in both bases $\boldsymbol{\mathcal{E}}$ and $\boldsymbol{\mathcal{P}}$. For that,  Eve splits the signal into two pulses of half intensity by using a 50/50 beam splitter on each mode and next performs  measurements of paired values for $N$ modes in the two bases, that is, $\{ (\mathcal{E}_{1},..,\mathcal{E}_{N}),(\mathcal{P}_{1},.. ,\mathcal{P}_{N})\}$.  Thus, if a  quantum state $\ket{\boldsymbol{a}}$ is sent, the probability that Eve measures an outcome $({\boldsymbol{\mathcal{E}}},{\boldsymbol{\mathcal{P}}})$ is given by:
\begin{equation*}
  Z({\boldsymbol{\mathcal{E}}},{\boldsymbol{\mathcal{P}}}) = \big\vert \braket{{\boldsymbol{\mathcal{E}}}\vert\dfrac{\boldsymbol{a}}{\sqrt{2}}}\big\vert^{2} \big\vert  \braket{ {\boldsymbol{\mathcal{P}}\vert \dfrac{\boldsymbol{a}}{\sqrt{2}}}}  \big\vert^{2}=
 \end{equation*}
  \begin{equation}\label{zeta}
  = (\dfrac{2}{\pi})^{N}\prod_{j=1}^{N}\exp{[-2(\mathcal{E}_{j}-\dfrac{a_{j}}{\sqrt{2}})^{2}]} \exp{[-2\mathcal{P}_{j}^{2}]}.
\end{equation}

Depending on the quadrature values, the most probable mutimode state  $\vert L\rangle = \ket{L_{1} ...L_{N}}$ can be  determined by using some criteria. In this work, we generalize, to a certain extent, the criteria proposed in  \cite{Namiki2003} for the case $N$$=$$1$  mode,  where  we have the states of the first quadrature $\vert L\rangle $$=$$\vert \pm a\rangle$  and the second quadrature $\vert L\rangle $$=$$\vert \pm ia\rangle$. Such a  criteria establishes that the most probable state is given by the measurement's greatest absolute value. Note that where the wrong basis is selected, the most probable measurement value is zero. Next,  Eve obtains a pair of values $\mathcal{E}_{1}$ and  $\mathcal{P}_{1}$ (optical field in the second quadrature) for each signal, therefore if $\mathcal{X}=\sqrt{\mathcal{E}_{1}^{2}}>\mathcal{Y}=\sqrt{\mathcal{P}_{1}^{2}}$ then a state $\vert L\rangle$$ =$$\vert$$ \pm $$a\rangle$ is obtained (the sign of  $\mathcal{E}_{1}$ determines which state has to be choosen). Otherwise, if $\mathcal{X}<\mathcal{Y}$ then  a state $\vert L\rangle$$=$$\vert $$ \pm $$ i a\rangle$ is obtained (the sign of  $\mathcal{P}_{1}$ determines which state has to be choosen).  Therefore, the  generalization to $N$ modes consists of defining the numbers $\mathcal{X}=\sqrt{\mathcal{E}_{1}^{2}+...+\mathcal{E}_{N}^{2} }$, $\mathcal{P}=\sqrt{\mathcal{P}_{1}^{2}+...+\mathcal{P}_{N}^{2} }$ and then an analogous criteria can be used, that is, the greatest modulus determines the Eve's choice of basis,  if $\mathcal{X}>\mathcal{Y}$ then a state $\vert L\rangle=\vert (-1)^{\mu_{1}} a \,....\, (-1)^{\mu_{N}} a\rangle$ ($\mu_{i}=\{0,1\}$) is choosen, and where the signs of $\mathcal{E}_{1}, ..., \mathcal{E}_{N}$ determines which state has to be selected. The same procedure is applied for the case $\mathcal{X}<\mathcal{Y}$.

On the other hand, in order to simplify the calculation,  we will use $N$ dimensional spherical coordinates \cite{Blumenson1960} for each conjugate space $\mathcal{E}/\mathcal{P}$ and  thus a  direct comparison of  the modulus of ${\boldsymbol{\mathcal{E}}}$ and $ {\boldsymbol{\mathcal{P}}}$ can be made. Therefore, the new variables can be written as follows
\begin{equation}
\label{xr}
    \mathcal{X} = d({\boldsymbol{\mathcal{E}}})=\sqrt{\mathcal{E}_{1}^{2}+...+\mathcal{E}_{N}^{2} }  \hspace{0.5cm} \mathcal{Y} = d({\boldsymbol{\mathcal{P}}})=\sqrt{\mathcal{P}_{1}^{2}+...+\mathcal{P}_{N}^{2} } 
\end{equation}
\begin{equation}
\label{thet}
    \theta_{j} = \arctan({\sqrt{d^{2}({\boldsymbol{\mathcal{E}}}) - \sum_{k=1}^{j}{\mathcal{E}^{2}_{k}}}\Big/ \ \mathcal{E}_{j}}) 
\end{equation}
\begin{equation}
\label{thetN}
    \theta_{N-1} = \arctan2(\mathcal{E}_{N}, \mathcal{E}_{N-1}) 
\end{equation}
\begin{equation}
\label{varph}
    \varphi_{j} = \arctan({\sqrt{d^{2}({\boldsymbol{\mathcal{P}}}) - \sum_{k=1}^{j}{\mathcal{P}^{2}_{k}}}\Big/ \ \mathcal{P}_{j}}) 
\end{equation}
\begin{equation}
\label{varphN}
    \varphi_{N-1} = \arctan2(\mathcal{P}_{N}, \mathcal{P}_{N-1}) 
\end{equation}
where the angles for the ${\boldsymbol{\mathcal{E}}}({\boldsymbol{\mathcal{P}}})$ variable are  $\theta_{1}(\varphi_{1}),..,\theta_{N-2}(\varphi_{N-2}) \in  [0,\pi ]$, $\theta_{N-1}(\varphi_{N-1}) \in [0,2\pi)$, %$d(\boldsymbol{x}) = \sqrt{x^{2}_{1}+..+x^{2}_{N}}$ is the Euclidean distance
 and $\arctan2$ is the two-argument arctangent function. As an example for $N=3$ in optical field space ${\boldsymbol{\mathcal{E}}}$, $\mathcal{X} = \sqrt{\mathcal{E}^{2}_1+\mathcal{E}^{2}_2+\mathcal{E}^{2}_3} = r$, $\theta_1 = \arctan({\sqrt{\mathcal{E}^{2}_{2}+\mathcal{E}^{2}_{3}}}/{\mathcal{E}_{1}} ) = \theta$ and $\theta_{2} =\arctan({\mathcal{E}_{3}}/{\mathcal{E}_{2}} ) = \varphi $ which are the triad of the wellknown variables $(r,\theta,\varphi)$ for spherical coordinates.
Finally, Eve's resending criteria for an state $\ket{L}$  in these generalized spherical coordinates is given by
\begin{equation}
     \ket{L}=
     	\begin{cases}
     		\ket{(-1)^{\mu_{1}} a_{1}}..\ket{(-1)^{\mu_{j}} a_{j}}..\ket{(-1)^{\mu_{N}} a_{N}} & \text{if } \mathcal{X} \geq \mathcal{Y} \\
     		\ket{(-1)^{\mu_{1}} i a_{1}}..\ket{(-1)^{\mu_{j}} ia_{j}}..\ket{(-1)^{\mu_{N}} ia_{N}}& \text{if } \mathcal{Y} \geq \mathcal{X}
     	\end{cases}
\end{equation}
with indexes for the optical field variable given by the following expressions
\begin{equation}
    \begin{cases}
	 \mu_{1} = \frac{1}{2}[1-s_{1}] & \text{with} \hspace{1.0mm}  s_{1} = \sign(\cos\theta_{N-1})   \\
         \mu_{2} = \frac{1}{2}[ 1-s_{2} ] &  \text{with} \hspace{1.0mm} s_{2} = \sign(\sin\theta_{N-1}) 
        \\ 
	  \mu_{j} =  \frac{1}{2}[1-s_{j}]  & \text{with} \hspace{1.0mm} s_{j} = \sign(\cos\theta_{j}) \hspace{1.0mm} \forall  j \in (3,N) 
    \end{cases}
\end{equation}
It represents a partition of the N-dimensional sphere into  $2^{N}$-orthants (for $N=3$ would be an octant). Eve identifies the orthant by looking  at the sign of each vector component. Because of this, the resending signal can be characterized by the probabilities of identifying properly the input state $p_{+...+}$, or only the basis but other state (e.g. $p_{-+...+}$,  $p_{+...-}$ and so on), or neither the state nor the basis $p_{\perp...\perp}$. As an example, by using Eq.(\ref{zeta}) the calculation of probability $p_{+...+}$  is as follows

\begin{equation}
 p_{+...+} = \int_{\mathcal{E}_{j}>0 \hspace{1mm}\forall j} [\int_{\mathcal{Y}>\mathcal{X}}{Z({\boldsymbol{\mathcal{E}}},{\boldsymbol{\mathcal{P}}}) d\mathcal{P}_{1}..d\mathcal{P}_{N}}] d\mathcal{E}_{1}..d\mathcal{E}_{N}
\end{equation}

Explicitly in generalized spherical coordinates the later expression results in:
\begin{align*}
    p_{+...+} = (\frac{2}{\pi})^N \exp[-\frac{|\boldsymbol{a}|^{2}}{2}] \int{dV_{N}}  \int_{0}^{\pi/2}{d\theta_{1}..d\theta_{N-1}\sin^{N-2}(\theta_1)..\sin(\theta_{N-2})} 
\end{align*}
\begin{equation}
    \cdot [\int_{0}^{\infty}{d\mathcal{X} [\int_{0}^{\mathcal{X}}{d\mathcal{Y} \,\mathcal{Y}^{N-1}\exp{[-\mathcal{Y}^{2}]}}] \mathcal{X}^{N-1}\exp{[-\mathcal{X}^{2}+2\sqrt{2}\mathcal{X}(\boldsymbol{a}\cdot\boldsymbol{u})]}}]
\end{equation}
where $dV_{N} =  \sin^{N-2}(\phi_1)\sin(\phi_{N-2})..d\phi_{1}..d\phi_{N-2} d\phi_{N-1} $ and the generalized spherical directional vector is  {$\boldsymbol{u} = (\sin(\theta_{1})...\sin(\theta_{N-1}),..., \cos(\theta_{1}))$. It can be  easily checked that in the low energy limit where $a_{i}<<1$ all these integrals converge to the limit $\frac{1}{2^{N+1}}$.

For sake of completeness, we write the general expresion for a general $\mathcal{E}$ basis measurement probability, where the above result is a particular case, that is, %and the $\mathcal{P}$ basis measurement probability:
\begin{equation}
 p_{s_{1}...s_{N}} = \int_{s_{j},\mathcal{E}_{j}>0 \hspace{1mm}\forall j} [\int_{\mathcal{Y}>\mathcal{X}}{Z({\boldsymbol{\mathcal{E}}},{\boldsymbol{\mathcal{P}}}) d\mathcal{P}_{1}..d\mathcal{P}_{N}}] d\mathcal{E}_{1}..d\mathcal{E}_{N}
\end{equation}
%\textcolor{red}{(AQUI NON ESTA CLARO SE $s_{j}$ e $\mathcal{E}_{j}$ son $>0$  ou só un deles , logo puxen coma)}
 where we make use of the sign operator $s_{j}=\pm$ and a positive $\mathcal{E}_{j}$ is employed for this calculation. Likewise, we write the $\mathcal{P}$ basis measurement probability  as follows
\begin{equation}
 p_{\perp...\perp} = \int_{\mathcal{P}_{j}>0 \hspace{1mm}\forall j} [\int_{\mathcal{Y}<\mathcal{X}}{Z({\boldsymbol{\mathcal{E}}},{\boldsymbol{\mathcal{P}}}) d\mathcal{E}_{1}..d\mathcal{E}_{N}}] d\mathcal{P}_{1}..d\mathcal{P}_{N}
\end{equation}
In short,  resended states are transformed by Eve's attack, for example, the state $\ket{a_{1}...a_{N}}$ is rewritten as:
\begin{equation*}
\nonumber \ket{a_{1}...a_{N}} \rightarrow p_{+...+} \ket{a_{1}...a_{N}} +...+ p_{-...-} \ket{-a_{1}...-a_{N}} + 
\end{equation*}
\begin{equation}\label{transf}
+p_{\perp...\perp} (\ket{ia_{1}...ia_{N}} +...+\ket{-ia_{1}...-ia_{N}})
\end{equation}
As a consequence of Eq.(\ref{transf}) the density operators from Bob's perspective (\ref{rho1}) and (\ref{rho2}) are transformed as follows
\begin{equation}
    \hat{\rho}^{'}_{1} = (\hspace{-0,25cm}\sum_{s_{1}...s_{N}=+,-}{p_{s_{1},...,{s_{N}}}})\hat{\rho}_{1} + (2^{N}p_{\perp...\perp})\hat{\rho}_{2}
\end{equation}
\begin{equation}
    \hat{\rho}^{'}_{2} = (2^{N}p_{\perp...\perp})\hat{\rho}_{1} + (\hspace{-0,25cm}\sum_{s_{1}...s_{N}=+,-}{p_{s_{1},...,{s_{N}}}}) \hat{\rho}_{2}.
\end{equation}
As made in reference \cite{Namiki2003} for $N=1$ to calculate Bob’s QBER in the presence of Eve, the $2^{N}$-PE has to be also  re-evaluated using Eq.(\ref{PE-N})  as follows
\begin{equation}
    P^{'}({\boldsymbol{\mathcal{E}_{0}}},\boldsymbol{a}) = (\hspace{-0,25cm}\sum_{s_{1}...s_{N}=+,-}{p_{s_{1},...,{s_{N}}}})P({\boldsymbol{\mathcal{E}_{0}}},\boldsymbol{a}) + 2^{N}p_{\perp...\perp}\prod_{i}{\erfc{\sqrt{2}}\mathcal{E}_{0,i}}.
\label{Pprima}
\end{equation}
By using  the above result % (\ref{Pprima}), 
the $2^{N}$-QBER under simultaneous measurement attack is defined, that is,
\begin{equation*}
q^{'} ({\boldsymbol{\mathcal{E}_{0}}},\boldsymbol{a}) = \frac{1}{P^{'}({\boldsymbol{\mathcal{E}_{0}}},\boldsymbol{a}) }\left [(\hspace{-0,05cm}\sum_{s_{1}...s_{N}=+,-}\hspace{-0,35cm}{p_{s_{1},...,{s_{N}}}}) (P(\boldsymbol{\mathcal{E}}_{0},\boldsymbol{a}) \right .
\end{equation*}
\begin{equation*}
- s_{1}\int^{s_{1}\infty}_{s_{1}\mathcal{E}_{0}}...\hspace{1mm}{s_{N}\int^{s_{N}\infty}_{s_{N}\mathcal{E}_{0}}{}\braket{\mathcal{E}_{1},...,\mathcal{E}_{N}|s_{1}a_{1},...,s_{N}a_{N}} d\mathcal{E}_{1}...d\mathcal{E}_{N}}  ) 
\end{equation*}
\begin{equation}
\left .+2^{N}p_{\perp...\perp} \int^{\infty}_{\mathcal{E}_{0}}{\braket{\mathcal{E}_{1},...,\mathcal{E}_{N}|\hat{\rho}_{2}|\mathcal{E}_{1},...,\mathcal{E}_{N}} d\mathcal{E}_{1}...d\mathcal{E}_{N}} \right ]
\end{equation}
Next, by solving the integrals the following expression for the $2^{N}$-QBER  is obtained
\begin{equation*}
q^{'} ({\boldsymbol{\mathcal{E}_{0}}},\boldsymbol{a}) = \frac{1}{2^{N }P^{'}({\boldsymbol{\mathcal{E}_{0}}},\boldsymbol{a}) } \left [ \right . (  \hspace{-0,25cm}\sum_{s_{1}...s_{N}=+,-}\hspace{-0,35cm}{p_{s_{1},...,{s_{N}}}}) P(\boldsymbol{\mathcal{E}}_{0},\boldsymbol{a})-
\end{equation*}
\begin{equation*}
\hspace{-0,25cm}-\sum_{s_{1}...s_{N}=+,-} \hspace{-0,35cm} p_{s_{1}...s_{N}}\erfc{\sqrt{2}(\mathcal{E}_{0,1}-s_{1}a_{1})}...
\end{equation*}
\begin{equation}
...\erfc{\sqrt{2}(\mathcal{E}_{0,N}-s_{N}a_{N})} +2^{N}(2^{N}-1)p_{\perp...\perp}\prod_{k=1}^{N}{\erfc{\sqrt{2}\mathcal{E}_{0,k}}} \left . \right ]
\end{equation}
This result can be rewritten as follows
\begin{align}
\nonumber q^{'} ({\boldsymbol{\mathcal{E}_{0}}},\boldsymbol{a}) = 1-\frac{1}{P^{'}({\boldsymbol{\mathcal{E}_{0}}},\boldsymbol{a})} [p_{\perp...\perp} \prod_{k=1}^{N}{\erfc{\sqrt{2}\mathcal{E}_{0,k}}}    \\ - \hspace{-0,25cm}\sum_{s_{1}...s_{N}=+,-}\hspace{-0,35cm}{p_{s_{1},...,{s_{N}}}}     \frac{\erfc{\sqrt{2}(\mathcal{E}_{0,1}-s_{1}a_{1})}...\erfc{\sqrt{2}(\mathcal{E}_{0,N}-s_{N}a_{N})}}{2^{N}} ]
\end{align}
which represents a QBER analogous to the IQBER given by Eq.(\ref{iqber}). The second term represents the mistakes measured by Bob due to Eve's wrong basis choice and the third term are the correct basis errors (but possible wrong state choice). Note that for a state $\ket{a_{1},...,a_{N}}$, if $p_{\perp...\perp} = 0 $ and  $p_{+...+} = 1$, that is, a free from attack situation then the Eq.(\ref{iqber}) is naturally recovered.
%\begin{align}
  %q^{'} ({\boldsymbol{\mathcal{E}_{0}}},\boldsymbol{a}) = \frac{P}{P^{'}} [(\hspace{-0,25cm}\sum_{s_{1}...s_{N}=+,-}\hspace{-0,35cm}{p_{s_{1},...,{s_{N}}})q^{} ({\boldsymbol{\mathcal{E}_{0}}},\boldsymbol{a}}) +\frac{2^{N}-1}{P}p_{\perp...\perp}\prod_{i}{\erfc{\sqrt{2}}\mathcal{E}_{0,i}}]
%\end{align}
In Fig.\ref{fig6}  a comparison is shown between the IQBER and the QBER in the presence of Eve for $N=$ 1, 2 and 3 modes for a typical threshold value.

\begin{figure*}[h]
\centering
\includegraphics[width=0.65\textwidth]{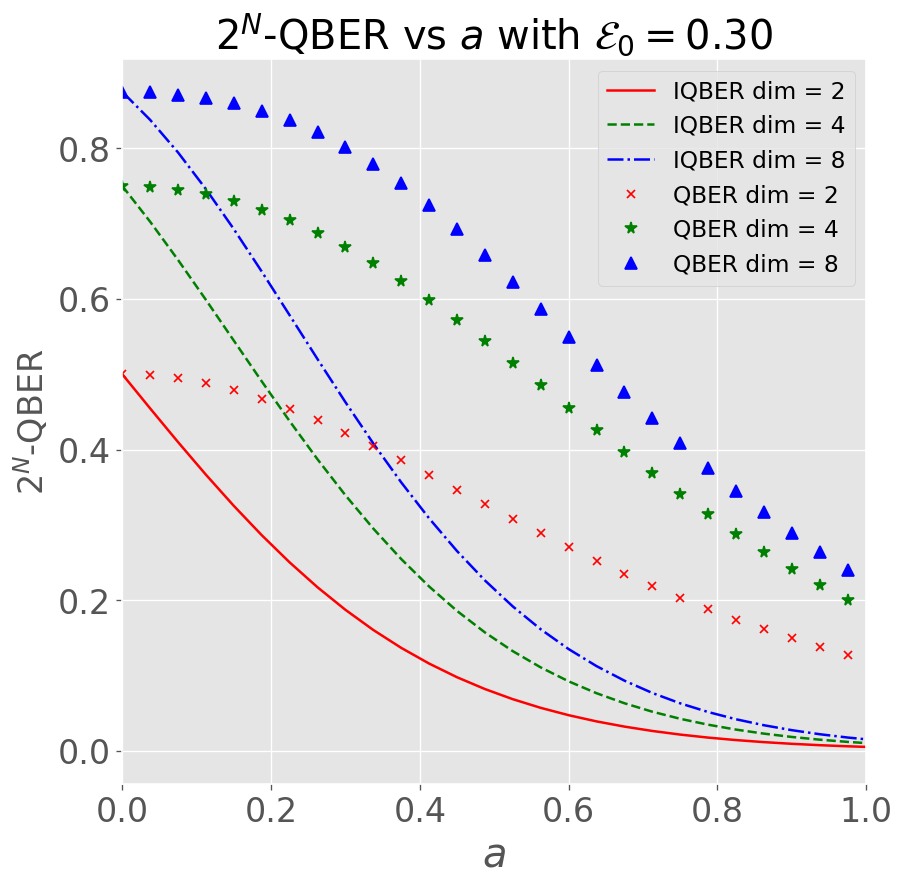}
\caption{A comparison between  the IQBER and the QBER for $N=$ 1, 2 and 3 modes,   as a function of the mean optical field $a$ in each mode, and with $\mathcal{E}_{0} = 0.30$. }
\label{fig6}
\end{figure*}

%%%%%%%%%%%%%%%%%%%%%%%%%%%%%%%%%%%%%%%%%%%%%%
%%%%%%%%%%%%%%%%%%%%%%%%%%%%%%%%%%%%%%%%%%%%%

\section{Secure key rate gain}

%%%%%%%%%%%%%%%%%%%%%%%%%%%%%%%%%%%%

%\subsection{Secure key gain rate model}
In order to estimate the performance of the protocol, a model which takes into account losses and the rate of interception of Eve is proposed. We must take into account,  on the one hand,  that the losses weaken  the  signal intensity and potentially it leads to an information leakage to Eve, and   on the other hand, if the rate of interception is too high, Eve is always detected and the channel communication is therefore interrupted. Overall, by taking into accout  the postselection efficiencies   $P$ (without Eve)   and $P'$ (with Eve) for an $\mathcal{E}$ basis measurement, the secure key gain  $G$ for a channel transmitance of $T$ and a rate of interception $\eta$ can be written as a function of the mutual information between Alice and Bob ($I_{AB}$) and the mutual information between Alice and Eve ($I_{AE}$) as follows
\begin{equation}
    G(\boldsymbol{\mathcal{E}_{0},\boldsymbol{a}},T,\eta) =\frac{1}{2}\{(\eta P^{'}+(1-\eta)P)I_{AB}(\boldsymbol{\mathcal{E}_{0}},\boldsymbol{a},T,\eta) -\eta P^{'} I_{AE}(\boldsymbol{a})\},
   \label{G}
\end{equation}
where the first term on the right side, that is,   the mutual information $I_{AB}$, is preceeded by a term of combination of postselections $P$ and $P'$  ponderated by the fraction of times $\eta$ that Eve attacks, and a prefactor $1/2$ that takes into account that only half of the times Alice and Bob will have a correct correspondence  between basis (quadratures),  that is,  the mutual information is reduced, in an effective form, by half \cite{Bourennane2002}. Note that Eve's attack on Alices states in the two bases provides a mutual information that is not limited, in principle, by the aforementioned basis correspondence. % this implies that the substracting term does not contain  \textcolor{red}{the ${1}/{2}$ prefactor,
 Accordingly,  we  obtain an upper limit for the mutual information  $I_{AE}$, and therefore a pessimistic quantitative assessment of the secure  key gain $G$, which  is entirely enough for our purpose of illustrating the security properties of the proposed protocol. % and phase.
 
The mutual information between Alice and Bob $I_{AB}$  is an extension of that one given in references \cite{Bourennane2002,Sheridan2010}, that is,
\begin{equation}\label{IAB}
      I_{AB}(\boldsymbol{\mathcal{E}_{0}},\boldsymbol{a},T,\eta) = H^{(\mathcal{N})}(\eta q^{'} + (1-\eta)q)
\end{equation}
where $H^{(\mathcal{N})}(p) = \log(\mathcal{N}) + (1-p)\log(1-p) + p \log(\frac{p}{\mathcal{N}-1})$, with $\mathcal{N}=2^{N}$, $p=\eta q^{'} + (1-\eta)q$  and $\log$ is the logarithm base 2. % \textcolor{magenta}{therefore we obtain
%\begin{equation*}
%I_{AB}(\boldsymbol{\mathcal{E}_{0}},\boldsymbol{a},T,\eta) = \log(\mathcal{N}) + [1-q-\eta(q'-q)]\log [1-q-\eta(q'-q)] + 
%\end{equation*}
%\begin{equation}\label{IAB}
%+ [q + \eta(q'-q) ] \log(\frac{ q + \eta(q'-q)   }{\mathcal{N}-1})
%\end{equation}
%}
Note that if there is no attack from Eve then $\eta = 0$ and  the secure key gain is given by % \textcolor{red}{it is obtained}
\begin{equation}
       G(\boldsymbol{\mathcal{E}_{0}},\boldsymbol{a},T,0) = \frac{1}{2}P \big(\log(\mathcal{N}) + (1-q)\log(1-q) +  q \log(\frac{q}{\mathcal{N}-1})\big)  
       \label{intri}
\end{equation}
that represents the intrinsic secure key gain due to inherent errors in the protocol which has no equivalent in DV-QKD. We must note that the proposed QKD system is  autocompensated, therefore all errors coming from polarization perturbations are neglected.

On the other hand, the mutual information between Alice and Eve $I_{AE}$ in Eq.(\ref{G}) is restricted only for the cases  where Bob makes a postselection, that is,  the prefactor  $P^{'}$ which in turn is multiplied by the fraction of times $\eta$ that Eve attacks. Besides, the Eve's attack is a simultaneous homodyne measurement in  bases  $\boldsymbol{\mathcal{E}}$ and $\boldsymbol{\mathcal{P}}$, therefore the probability to guess the basis is quite high and accordingly we can consider an overstimated mutual information   between Alice and Eve $I_{AE}$, that is, we assume that Eve always guesses the basis  equation {(assumption of getting the base right)} and therefore the correct quantum state with probability $p_{+...+}$ and the other quantum  states of the same basis with probabilities $p_{s_{1} ...s_{N}}$, where  $s_{1} ...s_{N}=\pm$.  Obviously, it provides an underestimated secure key gain, that is, it gives a greater amount of information to Eve which provides a more pesimistic secure key,  but it is enough to evaluate the security of the protocol;  
consequently, for this particular attack and assumption of getting the base right the overestimated mutual information between Alice and Eve $\tilde{I}_{AE}$    can be calculated as the difference between  {\em a priori} and {\em a posteriori} Shannon entropies \cite{Bourennane2002} and therefore the following expression is obtained 
\begin{equation}\label{I'AE}
{I_{AE}\leq \tilde{I}_{AE}=\log(\mathcal{N}) +\sum_{s_{1}...s_{N}=+,-} \Pi_{s_{1},...,{s_{N}}} \log(\Pi_{s_{1},...,{s_{N}}})}
\end{equation}
%I_{AE} = % ( \sum_{r_{1}...r_{N}=+,-}{p_{r_{1},...,{r_{N}}}} ) 
%\log(\mathcal{N}) +\sum_{s_{1}...s_{N}=+,-} {\Pi}_{s_{1},...,{s_{N}}}\log(\Pi_{s_{1},...,{s_{N}}}) 
%\begin{align}
% +\mathcal{N} p_{\perp...\perp} \left(\log(\mathcal{N})+  \sum_{s_{1}...s_{N}=+,-}{\tilde{\Pi}_{s_{1},...,{s_{N}}}}\log(\tilde{\Pi}_{s_{1},...,{s_{N}}})\right)
%\label{iae}
%\end{align}
where $\Pi_{s_{1},...,s_{N}}= p_{s_{1},...,s_{N}}/\sigma$, with  $\sigma=\displaystyle \sum_{s_{1}...s_{N}=+,-}\,\,p_{s_{1},...,{s_{N}}}$  the renormalized probabilities of Eve's correct guess of basis.   %and where $\tilde{\Pi}_{s_{1},...,{s_{N}}}= \tilde{p}_{s_{1},...,{s_{N}}}/\mathcal{N} p_{\perp...\perp}$ }are Eve's normalized  probabilities of guessing the state after measuring in the wrong basis, taking into account Alice's choice of basis and recalculating the results into the proper quadrature. These last probabilities are upper bounded because Eve can extract as much information as if she had guessed the basis, therefore  \textcolor{red}{$\tilde{\Pi}_{s_{1},...,{s_{N}}}\leq \Pi_{s_{1},...,{s_{N}}}$} and equation (\ref{iae}) is bounded by 
%where $\sum_{s_{1}...s_{N}=+,-}{p_{s_{1},...,{s_{N}}}} + N  p_{\perp...\perp} = 1 $ is employed. 

Next,  by using $I_{AB}$ and $\tilde{I}_{AE}$ given by Eq.(\ref{IAB}) and Eq.(\ref{I'AE}), respectively,  into Eq.(\ref{G}), we will obtain underestimated values of  $G$.
%\subsection{Simulation}
\begin{figure}[h]
     \centering
         \includegraphics[width=0.7\linewidth]{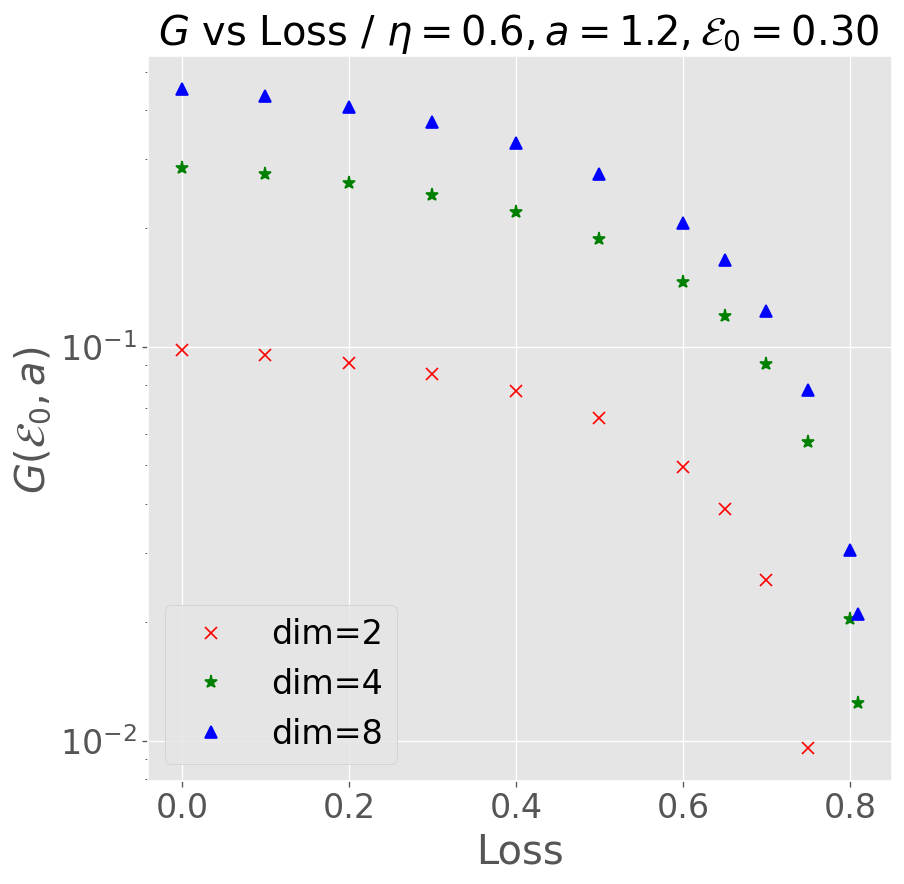}    
  %\begin{figure}[h]
% \centering
%\includegraphics[width=1.0\textwidth]{Keyrate}
 \caption{ Secure key gain for an intercept rate of $\eta = 0.60$ for  $\mathcal{N}$= 2, 4 and 8 dimensions with losses $\sqrt{1-T}$. }
\label{keyrate}
\end{figure}
The calculation of the  secure key gain given by Eq.(\ref{G}) has to be done by numerical techniques. In order to simplify the calculations, several reasonable choices of parameters are made.  First of all, same mean photon number $n$ and same losses $T$ are picked for each mode, therefore the same mean optical field $a_{i} = a$ and the same effective mean optical field  $a^{'} = \sqrt{T} a$ are chosen. Secondly, square type threshold frontiers are selected $\mathcal{E}_{0,i}=\mathcal{E}_{0}$, which is a reasonable constrain because there is no preferential choice between modes. This step allows us to simplify the numerical calculations afterwards. By setting a rate of Eve's attack $\eta$, for a given transmission $T$ (or Loss),  a value of gain $G(\sqrt{1-T})$  can be obtained for a pair of chosen values ($\mathcal{E}_{0}$,  $a$), although such value will be less than the  true secure key gain as commented above.

On the other hand, the increasing of secure key gain with higher dimensionality is shown in Fig.\ref{keyrate}, in particular a comparison between secure key gains for  dimensions $\mathcal{N}$= 2, 4 and 8  and   different levels of losses (or effective distance) and  for an intermediate  intercept rate $\eta=0.6$ (if a total attack is performed, that is,  $\eta = 1$, then the secure key gains are null).   Finally, note that the results  for $\mathcal{N}=2,4,8$ are always much better than for $N=1,2,3$ single channels, where the gain of $N$ single channels would be $N$ times the gain of one channel ($N=1$). %We must indicate that for high losses the underestimation of the gain gives rise to values of gain closer for $N>1$.
 \begin{table}[h]  
\centering      
\begin{tabular}{c c c c c}  
\hline                        
 \text{Loss}&\text{Distance}(km)& $G_{1}$&$G_{2}$ & $G_{3}$\\ [0.5ex] % inserts table %heading 
\hline  0.00 & 0 & 0.098 & 0.285 & 0.452 \\
0.10 & 2.29 & 0.095 & 0.275 & 0.432 \\
0.20 & 4.84 & 0.091 & 0.261 & 0.407 \\
0.30 & 7.74 & 0.085 & 0.243 & 0.373 \\
0.40 & 11.09 & 0.077 & 0.220 & 0.330 \\
0.50 & 15.05 & 0.066 & 0.188 & 0.275 \\
0.60 & 19.90 & 0.049 & 0.146 & 0.206 \\
0.65 & 22.80 & 0.039 & 0.120 & 0.166 \\
0.70 & 26.14 & 0.026 & 0.091 & 0.123 \\
0.75 & 30.10 & 0.010 & 0.057 & 0.078 \\
0.80 & 34.95 & - & 0.020 & 0.030 \\
0.81 & 36.06 & - & 0.012 & 0.021 \\ [1ex]   
\end{tabular} \label{table}  
\caption{ Numerical results of $G$  for $N=1,2, 3$ modes, that is, $G_{1}$, $G_{2}$  $G_{3}$.  %with theirs values of threshold $\mathcal{E}_{0}$ and mean optical field $a$ for several losses and $\eta = 0.5$.} 
Distances are given for a typical fiber losses of $0.2$dB/km for $1550$ nm.}
\end{table}

 In Table 1 the values of secure key gains are written for the calculations given in Fig.\ref{keyrate}. % It is shown that the thresholds $\mathcal{E}_{0}$ diminished with higher dimension and the mean photon number is almost the same for each mode. 
 Note that for a given value of losses (that is, distance) the secure key gain  is increased with the dimension of quantum states, or in other words, for a given value of secure key gain the  QKD distance is increased with the dimension of the quantum states.

\vspace{-0,5cm}
\section{Conclusions}
A high-dimensional continuous variable QKD protocol based on discrete modulation of $N$-dimensional product states of  weak coherent states has been presented. These multimode product states can be measured with standard balanced homodyne systems in optical communications. The protocol is non sensitive to phase drifts and  the polarization  perturbations are  compensated in a passive way (autocompensating) by using HWPs allowing to implement plug and play a high-dimensional QKD system. Different aspects of the security of the protocol have been studied under an intercept and resend attack based on a simultaneous attack and it has been shown that both  the QBER introduced by Eve and the secure key rate increase in a remarkable way with the dimension. It is crucial to highlight that the $2^{N}$ high dimensional cases provide much better results that the ones obtained by parallelization of $N$  two-dimensional (polarization)  lines. %for which we achieve better results.

\section*{Acknowledgments}

This work was supported in part by the MICIN, European Union NextGenerationEU under Grant PRTR-C17.I1, and in part by the Galician Regional Government through Planes Complementarios de I+D+I con las Comunidades Aut\'onomas in Quantum Communication. It was also funded by MCIU/ AEI / 10.13039/501100011033 / FEDER, UE under project PID2023-152607NB-I00.

%\appendix
%\section{Example Appendix Section}
%\label{app1}


%%       \cite{<label>} ==> [1]

%%
%Example citation, See \cite{lamport94}.


\bibliography{HD-CV-QKD.bib}
\bibliographystyle{elsarticle-num}




%\begin{thebibliography}{00}

%% For numbered reference style

%\bibitem{lamport94}
 % Leslie Lamport,
 % \textit{\LaTeX: a document preparation system},
 % Addison Wesley, Massachusetts,
 % 2nd edition,
  %1994.

%\end{thebibliography}

\end{document}